# Resistive superconducting transition and effects of atmospheric exposure in the intercalation superconductor $A_x(C_2H_8N_2)_yFe_{2-z}Se_2$ ($A$ = Li, Na)

**T. Hatakeda, T. Noji, S. Hosono, T. Kawamata, M. Kato, and Y. Koike**

*Department of Applied Physics, Tohoku University, 6-6-05 Aoba, Aramaki, Aoba-ku, Sendai 980-8579, Japan*

noji@teion.apph.tohoku.ac.jp

**Abstract**. We have succeeded in observing zero-resistivity in newly discovered intercalation superconductors $A_x(C_2H_8N_2)_yFe_{2-z}Se_2$ ($A$ = Li, Na) with $T_c$ = 45 K, using the sintered pellet samples. The electrical resistivity, $\rho$, in the normal state is metallic and $T_c^{onset}$, defined in the $\rho$ measurements, is as high as ~ 57 K. We have also investigated effects of the atmospheric exposure in $Li_x(C_2H_8N_2)_yFe_{2-z}Se_2$. It has been found that both the crystal structure and superconductivity are maintained at least up to several days, indicating this material is comparatively resistant to the atmospheric exposure.

## 1. Introduction

Recently, it has been found that the superconducting transition temperature, $T_c$, of the iron-based superconductor FeSe increases from 8 K to 40 - 46 K via intercalation of alkaline or alkaline-earth metals and ammonia or pyridine [1-5]. Moreover, we have succeeded in synthesizing new superconductors $A_x(C_2H_8N_2)_yFe_{2-z}Se_2$ ($A$ = Li, Na) with $T_c$ = 45 K via intercalation of alkaline metals and ethylenediamine (EDA), $C_2H_8N_2$ [6,7]. The drastic increase in $T_c$ via the intercalation has attracted great interest in relation to the mechanism and the application of the intercalation superconductors. For applications, zero-resistivity is important, but zero-resistivity is observed only in $Li_x(C_2H_8N_2)_yFe_{2-z}Se_2$ [6]. Moreover, $T_c^{zero}$, defined as the temperature where the electrical resistivity, $\rho$, reach zero, is much lower than $T_c$ obtained in the measurements of the magnetic susceptibility, $\chi$. For applications, the stability of the intercalation superconductors at the atmosphere is also important, but generally superconductors including alkaline metals are not so stable at the atmosphere [1].

In this paper, we have investigated the resistive superconducting transition of alkaline metals- and EDA-intercalated $A_x(C_2H_8N_2)_yFe_{2-z}Se_2$ ($A$ = Li, Na) and their stability at the atmosphere.

## 2. Experimental

Polycrystalline samples of FeSe were prepared by the solid-state reaction method. Starting materials were powders of Fe and Se, which were weighted stoichiometrically, mixed thoroughly and pressed into pellets. The pellets were sealed in an evacuated quartz tube and heated at 800℃ for 40 h. The obtained pellets of FeSe were pulverized into powder to be used as a host of the intercalation. Dissolved alkaline metal (Li or Na) in EDA was intercalated into the powdery FeSe as follows. An appropriate amount of the powdery FeSe was placed in a beaker filled with 0.2 M solution of pure alkaline metal dissolved in EDA. The reaction was carried out at 45℃ for 7 days. The product was washed with fresh EDA. All the process was performed in an argon-filled glove box. Both FeSe and the intercalated samples were characterized by powder x-ray diffraction using $CuK_\alpha$ radiation. For the intercalated samples, an airtight sample-holder was used. In order to detect the superconducting transition, $\chi$ was measured using a superconducting quantum interference device (SQUID) magnetometer. Measurements of $\rho$ were also carried out by the standard dc four-probe method. For the $\rho$ measurements, as-intercalated powdery samples were pressed into pellets. Then, the pellets were sintered at 170 or 200℃ for 20 - 30 h in an evacuated or argon-filled glass tube. In order to investigate effects of exposure to the atmosphere, as-intercalated samples, pelletized at room temperature, and sintered pellet samples were exposed to the atmosphere up to 7 days.

## 3. Results and discussion

Figure 1 shows powder x-ray diffraction patterns of as-intercalated samples of $A_x(C_2H_8N_2)_yFe_{2-z}Se_2$ ($A$ = Li, Na). The broad peak around $2\theta$ = 20° is due to the airtight sample-holder. Although there are unknown peaks, most of sharp Bragg peaks are due to the intercalation compound of $A_x(C_2H_8N_2)_yFe_{2-z}Se_2$ ($A$ = Li, Na) and the host compound of FeSe, so that they are able to be indexed based on the $ThCr_2Si_2$-type and PbO-type structures, respectively. Therefore, it is found that alkaline metal and EDA are intercalated into FeSe, while there remains a non-intercalated region of FeSe in the samples. The lattice constants of $Li_x(C_2H_8N_2)_yFe_{2-z}Se_2$ are calculated to be $a$ = 3.440(3) Å and $c$ = 20.81(3) Å. The $c$-axis length of $Na_x(C_2H_8N_2)_yFe_{2-z}Se_2$ is 21.9(1) Å. Taking into account our previous results that the intercalation of only lithium into Fe(Se,Te) has neither effect on the superconductivity nor crystal structure[8], it is concluded that not only lithium or sodium but also EDA has been intercalated.

Figure 2 displays the temperature dependence of $\chi$ in a magnetic field of 10 Oe on zero-field cooling (ZFC) and on field cooling (FC) for as-intercalated powdery samples consisting of $A_x(C_2H_8N_2)_yFe_{2-z}Se_2$ ($A$ = Li, Na) and FeSe. The superconducting transition is observed at 45 K. Taking into account the powder x-ray diffraction results, it is concluded that the superconducting transition is due to bulk superconductivity of $A_x(C_2H_8N_2)_yFe_{2-z}Se_2$ ($A$ = Li, Na). The positive value of $\chi$ will be due to magnetic impurities taken into the sample. It is noted that the second superconducting transition due to FeSe observed at 8 K in our previous paper [6] is not observed clearly in the present lithium-intercalated sample. This is reasonable, because powder x-ray diffraction peaks due to FeSe are strongly suppressed in the present lithium-intercalated sample, as shown in Fig. 1.

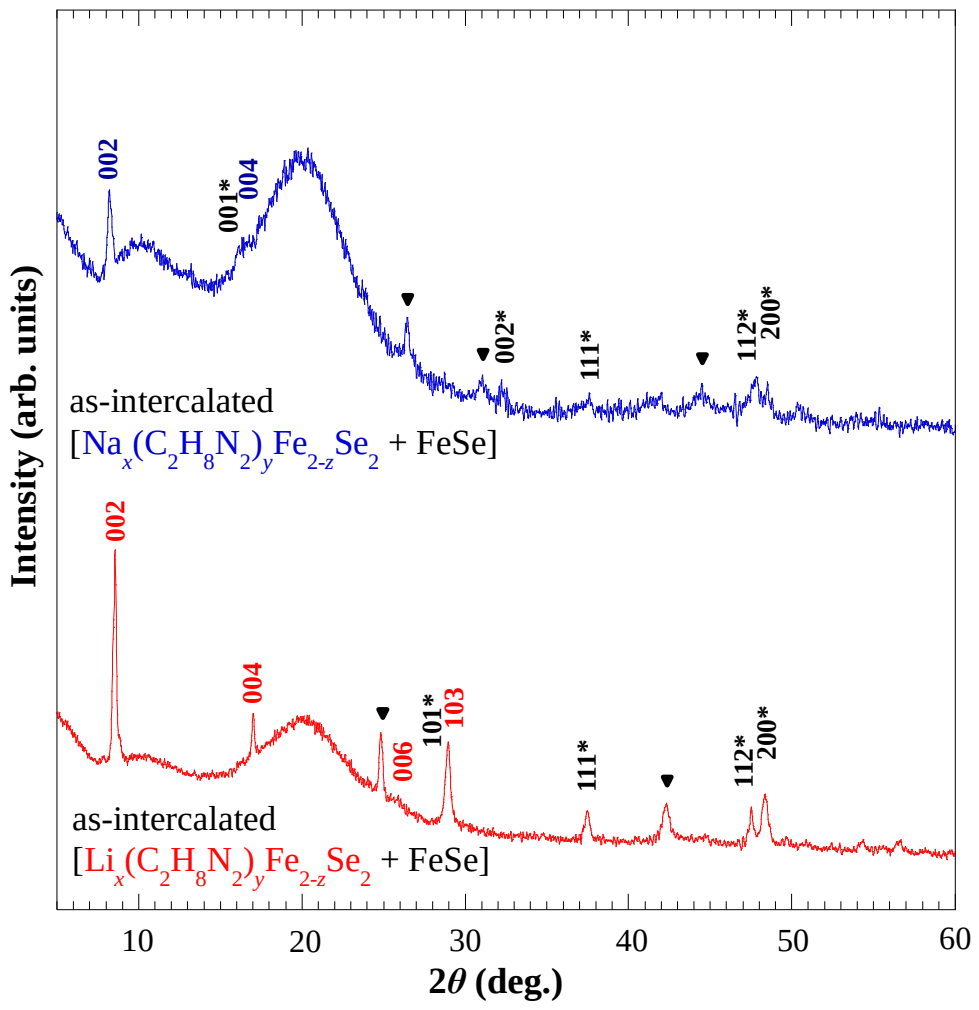

**Fig. 1.** Powder x-ray diffraction patterns of as-intercalated samples consisting of $A_x(C_2H_8N_2)_yFe_{2-z}Se_2$ ($A$ = Li, Na) and FeSe, using CuK$_\alpha$ radiation. Indexes without and with asterisk are based on the $ThCr_2Si_2$-type and PbO-type structures, respectively. Peaks marked by ▼ are unknown.

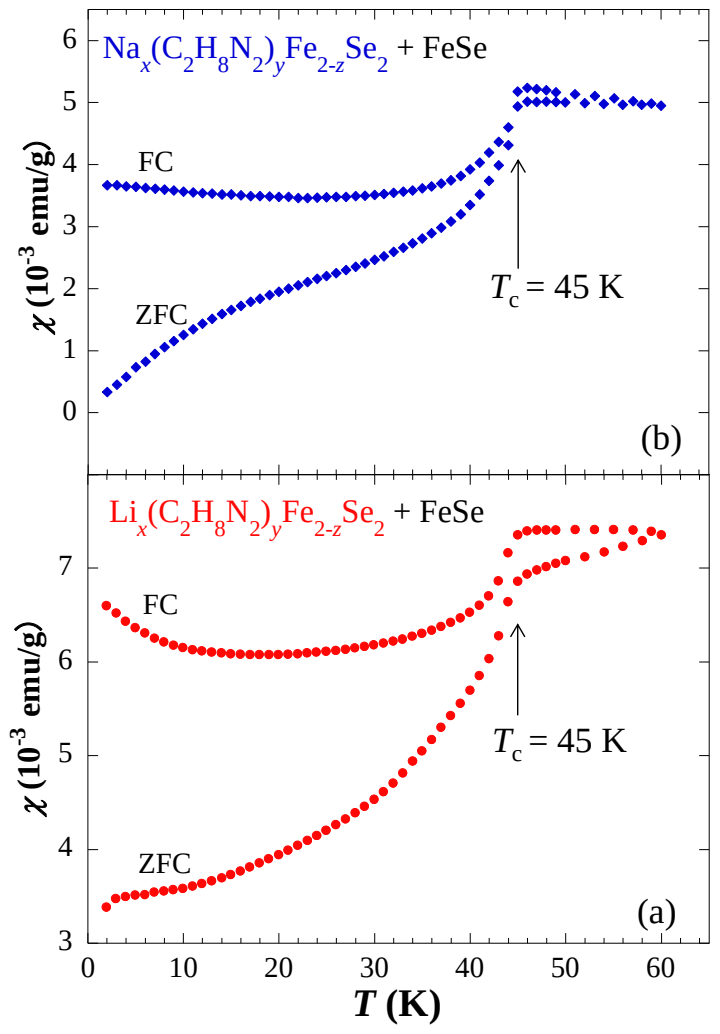

**Fig. 2.** Temperature dependence of the magnetic susceptibility, $\chi$, in a magnetic field of 10 Oe on zero-field cooling (ZFC) and field cooling (FC) for as-intercalated powdery samples consisting of $A_x(C_2H_8N_2)_yFe_{2-z}Se_2$ ((a) $A$ = Li, (b) $A$ = Na) and FeSe.

Figure 3 displays the temperature dependence of $\rho$ for several pellet samples. As shown in Fig. 3 (a), the as-intercalated sample consisting of $Li_x(C_2H_8N_2)_yFe_{2-z}Se_2$ and FeSe, pelletized at room temperature, exhibits the onset of the superconducting transition at 44 K, but zero-resistivity is not observed probably due to the insulating grain-boundary [6]. Sintered pellet samples consisting of $A_x(C_2H_8N_2)_yFe_{2-z}Se_2$ ($A$ = Li, Na) and FeSe, on the other hand, exhibit zero-resistivity. It is found that the lithium-intercalated pellet sample sintered at 170℃ shows a metallic temperature-dependence of $\rho$ and that the value of $\rho$ in the normal state is much smaller and $T_c^{zero}$ is much higher than those of the lithium-intercalated pellet sample sintered at 200℃, respectively. These results are understood to be due to the progress of the deintercalation of EDA with increasing sintering-temperature [7]. It is also found that $T_c^{mid}$, defined as the temperature where $\rho$ shows a half of the normal-state value, is ~ 42 K for both lithium- and sodium-intercalated pellet samples sintered at 170℃ and almost the same as $T_c$ = 45 K obtained from the $\chi$ measurements shown in Fig. 2. It is remarkable that $T_c^{onset}$, defined as the temperature where $\rho$ starts to decrease with decreasing temperature due to the superconducting transition, obtained from the $\rho$ measurements is as high as ~ 57 K and ~ 55 K for the lithium- and sodium-intercalated samples, respectively.

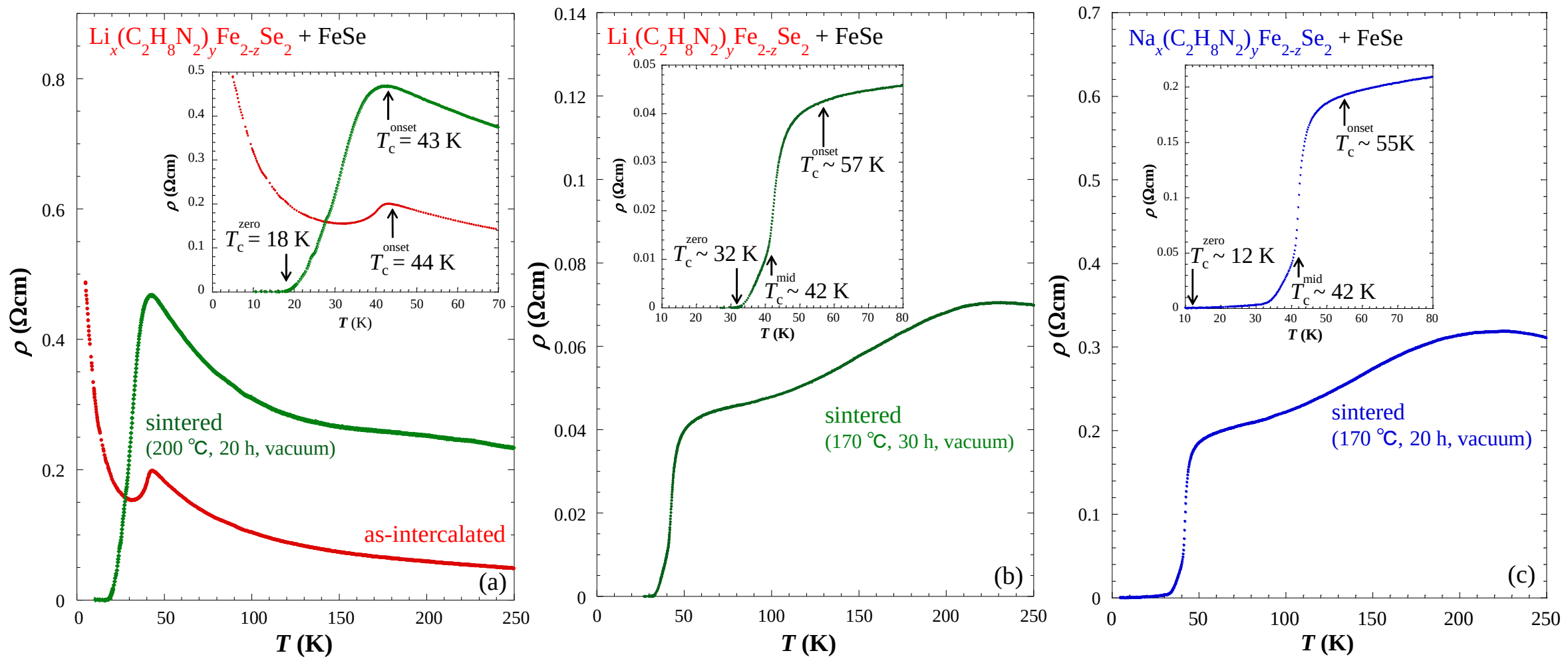

**Fig. 3.** Temperature dependence of the electrical resistivity, $\rho$, for (a) as-intercalated (pelletized at room temperature) and sintered (200℃, 20 h, vacuum) pellet samples consisting of $Li_x(C_2H_8N_2)_yFe_{2-z}Se_2$ and FeSe [6], (b) the sintered (170℃, 30 h, vacuum) pellet sample consisting of $Li_x(C_2H_8N_2)_yFe_{2-z}Se_2$ and FeSe, and (c) the sintered (170℃, 20 h, vacuum) pellet sample consisting of $Na_x(C_2H_8N_2)_yFe_{2-z}Se_2$ and FeSe. Insets show the temperature dependence of $\rho$ around $T_c$.

Figure 4 (a) shows the change of the x-ray diffraction pattern of the as-intercalated sample pelletized at room temperature, consisting of $Li_x(C_2H_8N_2)_yFe_{2-z}Se_2$ and FeSe, by exposure to the atmosphere. It is found that the diffraction peaks do not change so much even for the sample exposed for 7 days, indicating that the crystal structure of $Li_x(C_2H_8N_2)_yFe_{2-z}Se_2$ is maintained at the atmosphere at least up to 7days. Observing the enlarged diffraction patterns shown in Fig. 4 (b), it is found that the $c$-axis lengths of the samples exposed for 2 - 7 days are a little larger than that of the as-intercalated sample, as listed in Table I. The intercalants (lithium and EDA) may be relaxed structurally in the sample or lithium in the sample may react with something at the beginning of the exposure to the atmosphere. The change of the superconductivity of the as-intercalated sample pelletized at room temperature, consisting of $Li_x(C_2H_8N_2)_yFe_{2-z}Se_2$ and FeSe, by exposure to the atmosphere has been observed in the $\chi$ measurements, as shown in Fig. 5. It is found that both $T_c$ and the superconducting volume fraction gradually decrease by the atmospheric exposure. Furthermore, changes of $\rho$ and the resistive superconducting transition of the sintered (170℃, 30 h, Ar) pellet sample consisting of $Li_x(C_2H_8N_2)_yFe_{2-z}Se_2$ and FeSe by the atmospheric exposure have been observed, as shown in Fig. 6. It is found that the value of $\rho$ in the normal state gradually increases by the atmospheric exposure, while $T_c^{mid}$ = 34 - 36 K for all the samples. These results indicate that the superconductivity of $Li_x(C_2H_8N_2)_yFe_{2-z}Se_2$ is maintained at the atmosphere at least up to several days, though the superconducting volume fraction gradually decreases and the disorder at the grain boundary is gradually enhanced. Accordingly, it is concluded that $Li_x(C_2H_8N_2)_yFe_{2-z}Se_2$ is more resistant to the atmospheric exposure than the other intercalation superconductors such as $A_x(NH_3)_yFe_{2-z}Se_2$ ($A$ = alkaline and alkaline-earth metals) [1], so that $Li_x(C_2H_8N_2)_yFe_{2-z}Se_2$ is more suitable for applications.

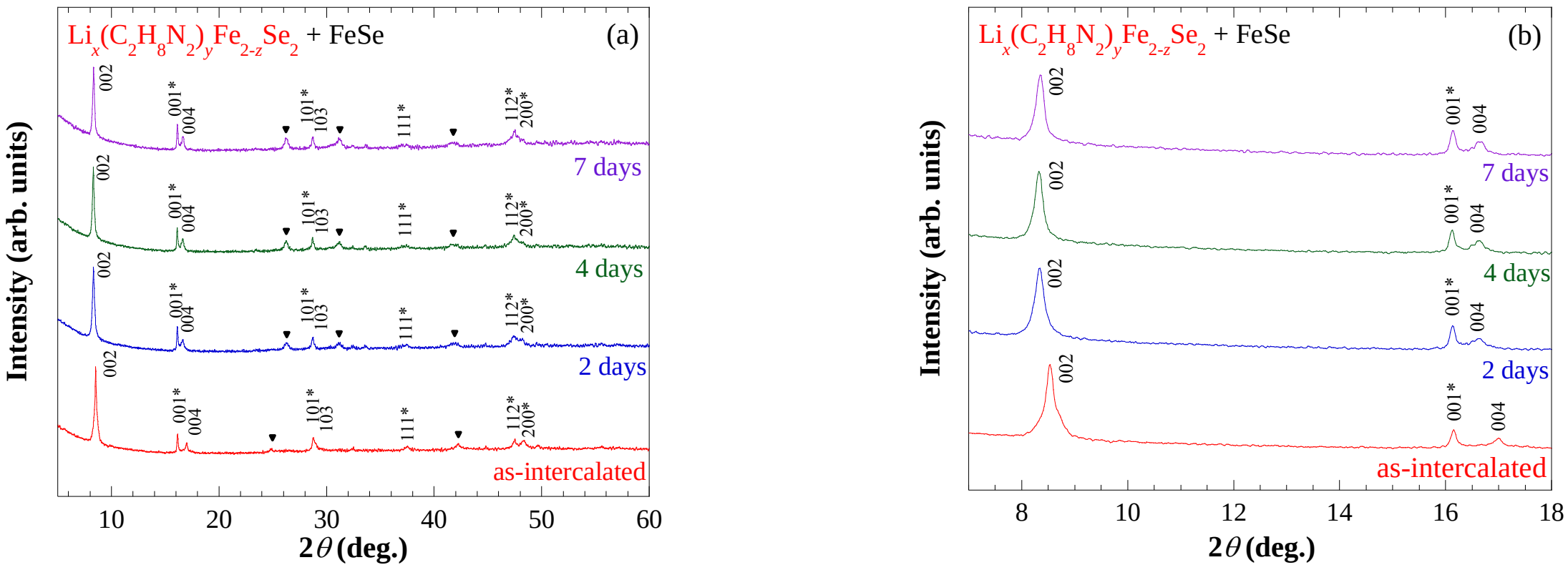

**Fig. 4.** (a) X-ray diffraction patterns of the as-intercalated sample pelletized at room temperature, consisting of $Li_x(C_2H_8N_2)_yFe_{2-z}Se_2$ and FeSe, and pellet samples exposed to the atmosphere for 2, 4 and 7 days using $CuK_\alpha$ radiation. (b) Enlarged x-ray diffraction patterns of (a) at $2\theta$ = 7 - 18°. Indexes without and with asterisk are based on the $ThCr_2Si_2$-type and PbO-type structures, respectively. Peaks marked by ▼ are unknown.

**Table 1.** Lattice constants *a* and *c* of $Li_x(C_2H_8N_2)_yFe_{2-z}Se_2$ for the as-intercalated sample pelletized at room temperature, consisting of $Li_x(C_2H_8N_2)_yFe_{2-z}Se_2$ and FeSe, and pellet samples exposed to the atmosphere for 2, 4 and 7 days.

| | *a* (Å) | *c* (Å) |
|---|---|---|
| $Li_x(C_2H_8N_2)_yFe_{2-z}Se_2$ (as-intercalated) | 3.447(3) | 20.85(4) |
| $Li_x(C_2H_8N_2)_yFe_{2-z}Se_2$ (2 days) | 3.426(2) | 21.33(3) |
| $Li_x(C_2H_8N_2)_yFe_{2-z}Se_2$ (4 days) | 3.435(2) | 21.32(2) |
| $Li_x(C_2H_8N_2)_yFe_{2-z}Se_2$ (7 days) | 3.433(2) | 21.30(3) |

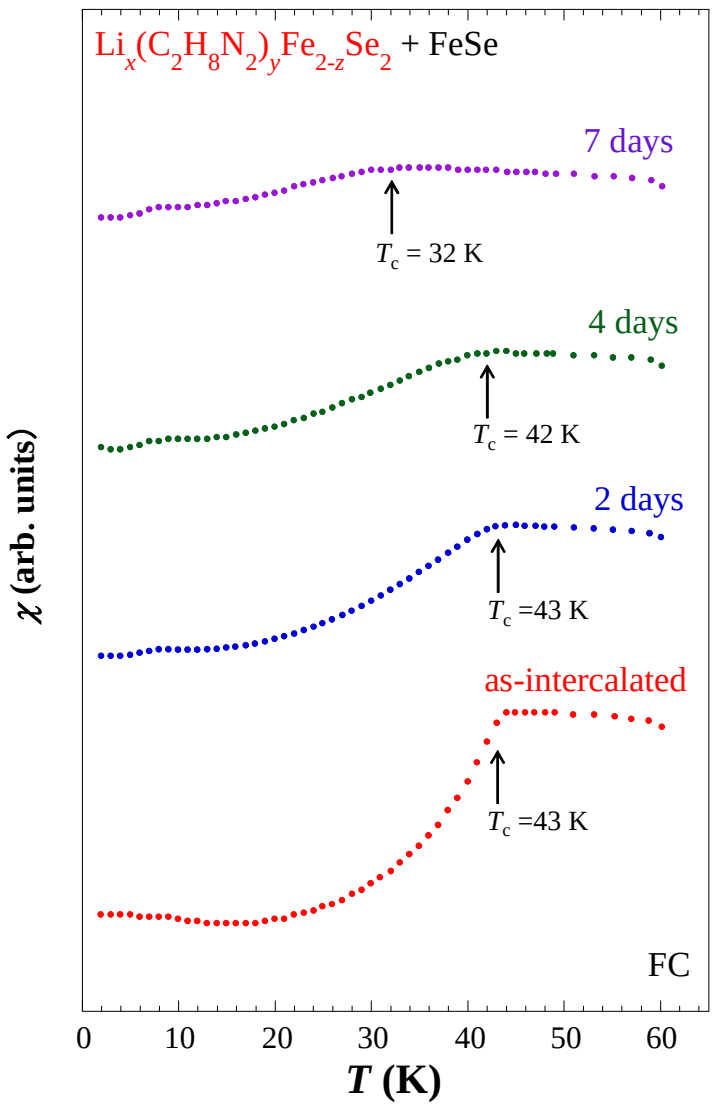

**Fig. 5.** Temperature dependence of the magnetic susceptibility, $\chi$, in a magnetic field of 10 Oe on field cooling (FC) for the as-intercalated sample pelletized at room temperature, consisting of $Li_x(C_2H_8N_2)_yFe_{2-z}Se_2$ and FeSe, and pellet samples exposed to the atmosphere for 2, 4 and 7 days.

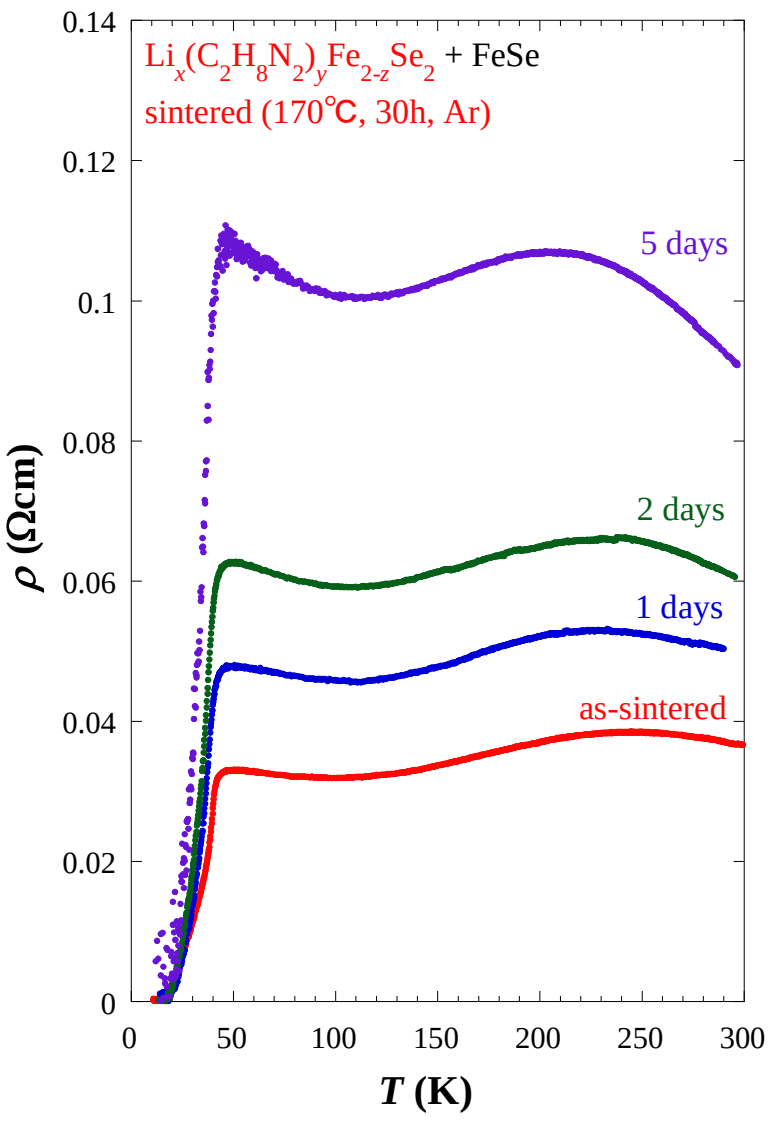

**Fig. 6.** Temperature dependence of the electrical resistivity, $\rho$, for the as-sintered (170℃, 30 h, Ar) pellet sample, consisting of $Li_x(C_2H_8N_2)_yFe_{2-z}Se_2$ and FeSe, and sintered (170℃, 30 h, Ar) pellet samples exposed to the atmosphere for 1, 2 and 5 days.

## 4. Summary

We have succeeded in observing zero-resistivity in our newly discovered intercalation superconductors $A_x(C_2H_8N_2)_yFe_{2-z}Se_2$ ($A$ = Li, Na) with $T_c$ = 45 K, using pellet samples sintered at 170℃ in an evacuated or argon-filled glass tube. It has been found that the sintered pellet samples show a metallic temperature-dependence of $\rho$ and that $T_c^{zero}$ is ~ 32 K and ~ 12 K for $A$ = Li and Na, respectively. For both $A$ = Li and Na, $T_c^{mid}$ is ~ 42 K and almost the same as $T_c$ obtained from the $\chi$ measurements. Moreover, $T_c^{onset}$ is as high as ~ 57 K and ~ 55 K for $A$ = Li and Na, respectively.

In order to investigate the stability of the lithium- and EDA-intercalated pellet samples at the atmosphere, they have been exposed to the atmosphere up to 7 days. From the x-ray diffraction analysis, it has been found that the crystal structure of $Li_x(C_2H_8N_2)_yFe_{2-z}Se_2$ is maintained at the atmosphere at least up to 7 days, though the $c$-axis lengths of the samples exposed for 2 - 7 days are a little larger than that of the as-intercalated sample. From the $\chi$ measurements, it has been found that the superconductivity is maintained at least up to 7 days, though both $T_c$ and the superconducting volume fraction gradually decrease. From the $\rho$ measurements of the sintered pellet samples, moreover, it has been found that zero-resistivity is maintained at least up to 5 days, though the value of $\rho$ in the normal state gradually increases. In conclusion, $Li_x(C_2H_8N_2)_yFe_{2-z}Se_2$ is comparatively resistant to the atmospheric exposure and comparatively suitable for applications.

**Acknowledgments**

One of the authors (T. H.) is indebted to the Motizuki Foundation for the support of his travel expenses to Buenos Aires to present this work at the 27th International Conference on Low Temperature Physics.

## References

[1] T. P. Ying, X. L. Chen, G. Wang, S. F. Jin, T. T. Zhou, X. F. Lai, H. Zhang, W. Y. Wang: Sci. Rep. **2** (2012) 426.
[2] E.-W. Scheidt, V. R. Hathwar, D. Schmitz, A. Dunbar, W. Scherer, F. Mayr, V. Tsurkan, J. Deisenhofer, A. Loidl: Eur. Phys. J. B **85** (2012) 279.
[3] M. Burrard-Lucas, D. G. Free, S. J. Sedlmaier, J. D. Wright, S. J. Cassidy, Y. Hara, A. J. Corkett, T. Lancaster, P. J. Baker, S. J. Blundell, S. J. Clarke: Nat. Mater. **12** (2013) 15.
[4] A. Krzton-Maziopa, E. V. Pomjakushina, V. Y. Pomjakushin, F. Rohr, A. Schilling, K. Conder: J. Phys.: Condens. Matter **24** (2012) 382202.
[5] L. Zheng, M. Izumi, Y. Sakai, R. Eguchi, H. Goto, Y. Takabayashi, T. Kambe, T. Onji, S. Araki, T. C. Kobayashi, J. Kim, A. Fujiwara and Y. Kubozono : Phys. Rev. B **88** (2013) 094521.
[6] T. Hatakeda, T. Noji, T. Kawamata, M. Kato and Y. Koike: J. Phys. Soc. Jpn. **82** (2013) 123705.
[7] T. Noji, T. Hatakeda, S. Hosono, T. Kawamata, M. Kato, Y. Koike: Physica C (in press)
[8] H. Abe, T. Noji, M. Kato, Y. Koike: Physica C **470** (2010) S487.